\documentclass{iopconfser}

\usepackage{amsmath}
\usepackage{lineno}
\usepackage{graphicx}
\usepackage{booktabs}
\usepackage{xcolor}
\usepackage{cite}
\usepackage{parskip}
\usepackage{url}

\newcommand{\un}[1]{\ensuremath{\,\mathrm{#1}}}
\newcommand{\chem}[1]{\ensuremath{\mathrm{#1}}}
\newcommand{\degC}[0]{\ensuremath{^\circ\mathrm{C}}}

\newcommand{\harvardcite}{} 
\newcommand{\citeasnoun}[1]{\cite{#1}}

\begin{document}


\title{The ESRF dark-field x-ray microscope at ID03}

\author{H.~Isern$^{1}$, 
    T.~Brochard$^{1}$,
    T.~Dufrane$^{1}$,
    P.~Brumund$^{1}$,
    E.~Papillon$^{1}$,
    D.~Scortani$^{1}$,
    R.~Hino$^{1}$,
    C.~Yildirim$^{1}$, 
    R.~Rodriguez Lamas$^{1}$,
    Y. Li$^1$,
    M. Sarkis $^1$,
    and C.~Detlefs$^{1}$
}

\affil{$^1$ European Synchrotron Radiation Facility, 38043 Grenoble, France}



\begin{abstract}
Dark Field X-ray Microscopy (DFXM) is a full-field imaging technique for non-destructive 3D mapping of orientation and strain in crystalline elements. The new DFXM beamline at ID03, developed as part of the ESRF Phase II Upgrade Project (EBSL2), was designed to provide cutting-edge capabilities for studying embedded microstructures. The project relocated and upgraded the end station from ID06-HXM to ID03, integrating new X-ray optics, radiation hutches, and a source device optimized for this advanced technique. Notable improvements include a near-field camera, a new goniometer, and a high-resolution far-field camera. The conceptual design was completed in September 2019, followed by the technical design in March 2021, with first users welcomed in April 2024. Building on the success of the original instrument, the ID03 beamline offers enhanced multi-scale and multi-modal mapping of microstructures with high resolution, enabling in-situ exploration of complex phenomena. Applications range from strain and orientation mapping in metals to studies of functional materials, semiconductors, biominerals, and energy systems.

\end{abstract}

\setlength{\parindent}{0pt}
\section{Introduction}

The need to study crystalline microstructures is persistent throughout materials science \cite{Callister2000}. Hierarchically organized crystalline structures, spanning up to 6 orders of magnitude in length scale, are ubiquitous in technological materials (both structural and functional), biominerals, geological materials and many others.
The microscopic crystal structure (e.g.~grains or domains) and the atomic-scale defect networks embedded therein determine many of the macroscopic physical and mechanical properties of these materials. Dark-field X-ray microscopy (DFXM) has been developed over the last 10 years to study the crystalline microstructure of materials with sub-micrometer spatial resolution \cite{Roth2015,Simons2015,Simons2016,Kutsal2019,yildirim2020}.
By combining hard x-ray diffraction and full-field imaging, it allows non-destructive mapping of the structure, orientation and strain of deeply embedded crystalline elements in three dimensions \cite{Poulsen2017,Poulsen2018,Jakobsen2019, yildirim2023extensive}.
An objective lens, in most cases a compound refractive lens (CRL), is placed between the sample and a high-resolution 2D detector, forming a magnified real-space image.

A prototype dark-field X-ray microscope was first installed on beamline ID06-HXM at the European Synchrotron Radiation Facility (ESRF, Grenoble, France) \cite{Kutsal2019}. This instrument operated from 2015 until 2022, when its end station was dismantled to serve as the foundation for the new dedicated dark-field X-ray microscopy beamline ID03 presented here, which operated with a user program during its final two years. In this article, we present the instrumentation of the new microscope at ID03, detailing all relevant hardware components.


\subsection{Principle of DFXM}

\begin{figure}
    \centering
    \includegraphics[width=0.8\linewidth]{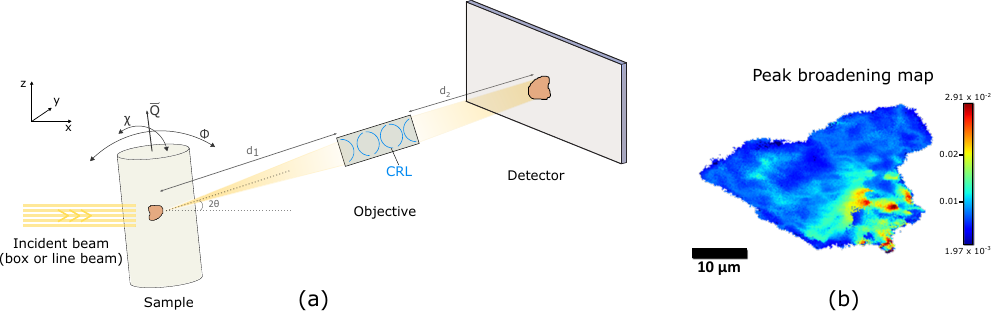}
    \vspace{-5mm}
    \caption{(a) Schematic of dark-field X-ray microscopy, showing monochromatic X-rays illuminating a crystalline sample. The diffracted beam is focused by a compound refractive lens (CRL) onto a 2D detector. Distances $d_1$, $d_2$, and $f_N$ represent the sample-CRL entry, CRL exit-to-image plane, and focal length, respectively. Orientation and strain are mapped by scanning tilt ($\chi$, $\phi$) and scattering ($2\theta$) angles, with sample rotation around the scattering vector $\vec{Q}$ providing additional projections. (b) Example DFXM Peak broadening map of a recrystallized ferrite grain generated through Gaussian fitting at each pixel.}
    \vspace{-2mm}

    \label{fig:geometry}
\end{figure}

The geometry and operational principle of dark-field x-ray microscopy (see figure~\ref{fig:geometry}) is akin to dark-field transmission electron microscopy (TEM): downstream of the sample, the diffracted beam passes through an x-ray objective lens, which creates a magnified image of a specific region of interest with contrast from local variations in lattice symmetry, orientation and strain \cite{Poulsen2017}. A defining feature of the dark-field x-ray microscope is the x-ray objective lens. The focal length and numerical aperture of the x-ray objective lens can be reconfigured to adjust the field-of-view, magnification and spatial resolution according to specific experimental requirements. So far compound refractive x-ray lenses (CRLs) made of Beryllium \cite{Snigirev1996} have been primarily used. 
Like the TEM, the dark-field x-ray microscope can be operated in a variety of modes. Most experiments so far have used either the \emph{projection} or \emph{section topography} configuration.

\begin{itemize}
\item In \emph{projection} configuration, the sample is illuminated by a ``box beam'' that is larger than the field-of-view. The recorded image is a projection of the sample along the diffracted beam direction, losing depth information.

\item In \emph{section topography} configuration, the sample is illuminated by a narrow line-beam \cite{Stoehr2015}, created by a one-dimensionally-focusing condenser. The illuminated layer is then imaged at an oblique angle.

\end{itemize}

User can switch back and forth between these configurations by inserting and removing the condenser lens.
Experiments typically involve the use of a succession of modalities which may include:

\begin{itemize}
\item \emph{Rocking curve imaging in section topography:}  The dependence of the intensity on the Bragg angle (rocking scan) is analyzed pixel by pixel. It is possible to combine rocking curve imaging in magnified (using the objective lens) and non-magnified (using the near-field camera, see below) modes without unmounting the sample \cite{Tran2021}.

\item \emph{Mosaicity scans:} By systematically varying the sample tilts, $\chi$ and $\phi$, a spatially resolved local pole figure can be acquired \cite{Poulsen2017,Garriga2023}.

\item \emph{Strain scans:} By scanning longitudinally ($\theta$--$2\theta$-scan) the strain component along the scattering vector is imaged \cite{Poulsen2017}. Typically, this is combined with a rocking or a mosaicity scan or with integration over the rocking profile at each $2\theta$ setting. As an alternative, strain mapping may be performed in the back focal plane \cite{Poulsen2018}.

\item \emph{Reciprocal space maps:} A high resolution reciprocal space map of the illuminated volume  is available either in the back focal plane \cite{Poulsen2018} or in the far-field regime (without the objective lens).   
\end{itemize}

3D mapping can be performed in two ways: The first method is to stack \emph{section topography} layers of the kind described above. This is performed by translating the sample through the planar beam in small increments. A second, faster method is \emph{magnified topo-tomography}. Here, projections of the sample are acquired while the sample is rotated about the scattering vector \cite{Ludwig2001,Poulsen2017}, and a 3D representation is reconstructed using tomographic principles. Again, data can be taken in magnified and non-magnified mode. Experimental protocols and reconstruction codes for magnified topo-tomography are currently under development.


\subsection{Complementary techniques}

The ultimate aim of the beamline is to enable multiscale 3D studies of crystalline materials, i.e.~to offer comprehensive full-field imaging of the phases, grains, domains, stress-fields, and defects within 
of mm-sized samples with the possibility to study local embedded features with a variable resolution from 10\un{\mu m} down to 30\un{nm}. 

Such a multi-length-scale approach requires the implementation of several x-ray techniques in one instrument. 
For the success of DFXM studies, in particular, it is essential that the sample can be characterized in-situ, e.g.~that grains or volumes of interest embedded within a larger sample can be located before proceeding to DFXM, providing intergranular context. The beamline is specifically designed to fulfill these requirements.  Techniques available include:
DFXM,
bright field microscopy  \cite{Lengeler1999,Falch2016b} and tomography,
classical diffraction topography and section topography, reciprocal space mapping,
3DXRD \cite{Poulsen2001,Schmidt2004,Jakobsen2006,Hefferan2012,osti_1476076} and diffraction contrast tomography (DCT) \cite{King2008,Ludwig2009}.


\subsection{Pink beam operation}

DFXM is a notoriously photon-starved technique. Many users of ID06-HXM have therefore requested pink beam operation --- the wider band width afforded by a multilayer monochromator instead of a classical crystal monochromator results in approximately $100\times$ higher flux. 
This comes at the cost of chromatic aberrations \cite{Falch2016a,Falch2018a}. 
Furthermore, complementary techniques could benefit from relaxing the resolution function \cite{Poulsen2017}, e.g.~texture tomography \cite{Grunewald2016,Frewein2024,Carlsen2024}.


\section{Beamline configuration and operating modes}

The beamline was built in the ESRF ID sector 03. 
It comprises one optics hutch (OH1) and one experiments hutch (EH1). 
OH1 is 13.3\un{m} long, 3.5\un{m} wide and 3.4\un{m} high. The air temperature is regulated to $\pm 1\degC$.
EH1 hosts the end-station.
It is 14.0\un{m} long, 4.8\un{m} wide and 4.0\un{m} high. 
The walls, ceiling and floor are thermally insulated, and the air temperature is regulated to $\pm 0.1\degC$.
A lead shielded, 8.8\un{m} long UHV pipe connects the two hutches.


\begin{figure}
    \centering
    \includegraphics[width=0.8\linewidth]{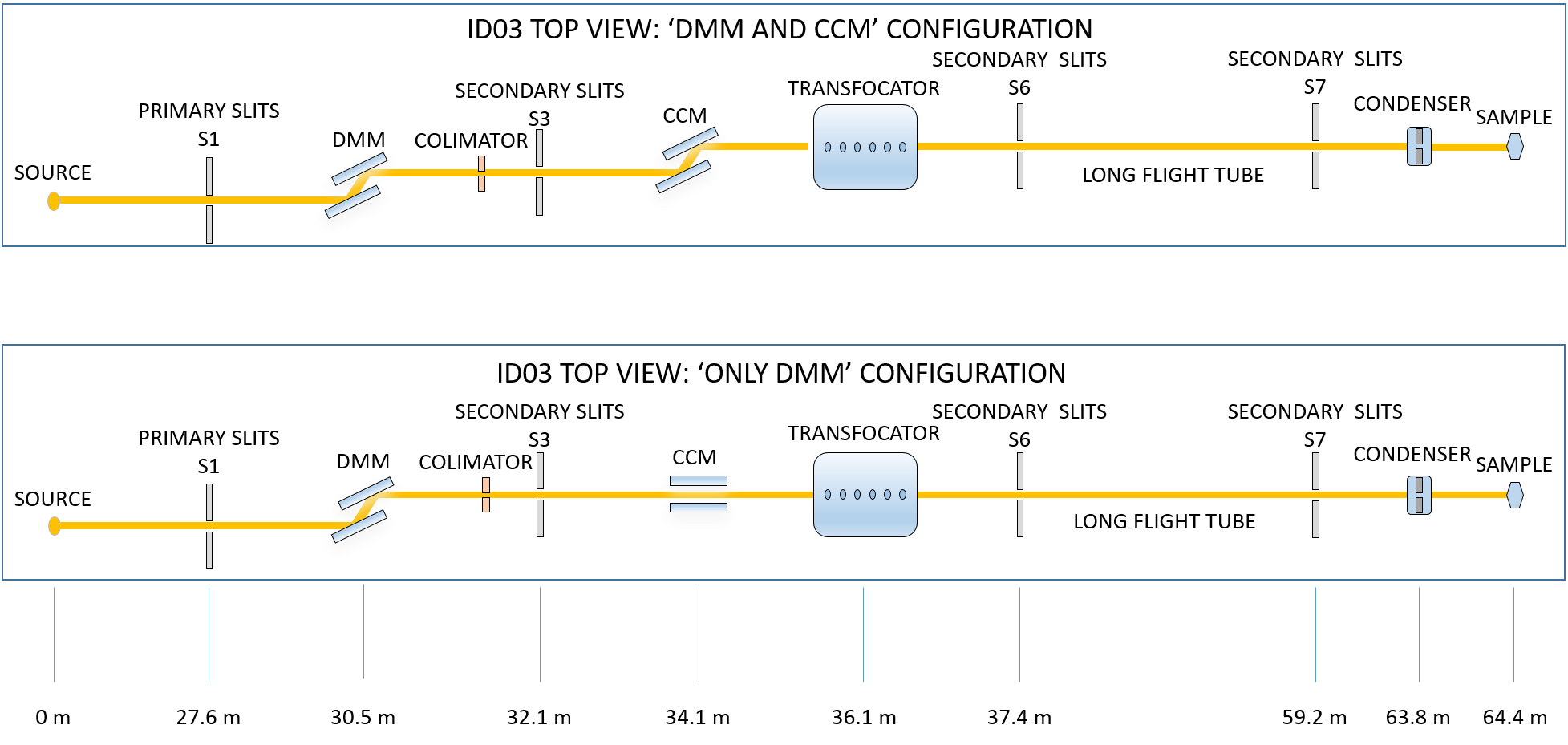}
    \caption{Top view scheme of two optics configurations: (top) using the Double Multilayer Monochromator (DMM), Channel-cut Crystal Monochromator (CCM) and Transfocator, and (bottom) using Double Multilayer Monochromator (DMM) and Transfocator only, without Channel-cut Crystal Monochromator (CCM)}
    \label{fig:config12}
\end{figure}

The beamline can be operated in ``pink'' or ``monochromatic'' mode. 
In both cases, the first optical elements (primary slits, white beam viewer, high heat load attenuators and double multilayer monochromator) are always in the beam path, see Fig.~\ref{fig:config12}. In pink beam operation, the beam passes through the transfocator and into the experiments hutch. In monochromatic operation, a channel cut crystal monochromator (symmetric Si(111)) is inserted in the beam path, offsetting the beam by $\approx 8\un{mm}$ to port ($+y$).
Again, the beam passes through the transfocator and into the experiments hutch.
The downstream optical elements are mounted on translation stages such that they can be used in both beam modes.


\section{Description of beamline components}

\subsection{Source and Front End}
The x-ray beam source is a 2\un{m} long CPMU with a period of 16.4\un{mm}, designed for a minimum gap of 4\un{mm}.
Pending machine upgrades, however, the gap is currently limited to $\geq 5\un{mm}$. For this, a liquid nitrogen sub-cooler system and a phase separator are installed near the ID03 straight section. Apertures in the Front End limit the beam size to $2\un{mm}(h) \times 1\un{mm}(v)$ in order to reduce the heat load on the downstream optical elements --- with the undulator fully closed to 4\un{mm}, this heat load is 3.6\un{kW}.
All optical elements are designed for the maximum heat load at gap 4\un{mm}.


\subsection{Optics hutch EH1}

The beamline optics are arranged in ``blocks''. 
Each block is mounted on a granite chassis that is bolted onto a steel plate, which is in turn glued to the floor. 
Granite was chosen for its low thermal expansion, high density, high stiffness, and high internal damping.


\subsubsection{Block 1: White beam} --- Optics Block 1 contains the following elements:

\paragraph{Primary Slits:}
This new generation of high power primary slits was developed at the ESRF 
for the high power output of CPMUs such as the one installed in the ID03 beamline.
They consist of two 329\un{mm} long, ``L''-shaped, water-cooled absorbers made out of CuCr1Zr (Copper Chromium Zirconium) alloy. The angle of incidence is 0.55$^\circ$ to spread the heat load.
Offset and gap can be adjusted by moving the absorbers parallel or antiparallel in the horizontal ($y$) and vertical ($z$) directions.

\paragraph{High-power white beam attenuators:}

\begin{table}
\centering
\begin{tabular}{cccc}
\toprule
Actuator & \#1 & \#2 & \#3 \\
         & (upstream) & (middle) & (downstream) \\
\midrule
Position 1 & 300\un{\mu m} & 300\un{\mu m} + 900 \un{\mu m} & 300\un{\mu m} +900\un{\mu m}\\
Position 2 & 300\un{\mu m} + 300\un{\mu m} & (empty) & (empty)\\
Position 3 & 300\un{\mu m} + 600\un{\mu m} & (empty) & (empty)\\
\bottomrule
\end{tabular}
\caption{Diamond filters of the high power white beam attenuator. Each of the 3 independent actuators can be equipped with up to 6 filters, 3 on the upstream side and 3 on the downstream side.}
\label{tab:hpatt}
\end{table}

The model used is a standard ESRF design. 
The main purpose of this device is to reduce the heat-load on the Double Multilayer Monochromator (DMM) and experiments using the pink beam mode. 
Furthermore, it absorbs low-energy photons that may pass through the DMM by total reflection when used at low angles (i.e.~high photon energies).
The device has three independent actuators, which are each equipped with water-cooled CVD diamond filters (see Table~\ref{tab:hpatt}).

\paragraph{White beam intensity and profile monitor:}


The ``beam viewer'' is also a standard ESRF design. 
It allows imaging the white beam via a diamond crystal used as scintillator and a camera. 
Furthermore a photodiode offers another means of measuring the beam intensity. 



\subsubsection{Block 2: Double Multilayer Monochromator (DMM)} --- Positioned at 30.5\un{m} from the source, the DMM is used as ``pink'' beam monochromator to reduce the heat load on the downstream optics and for high harmonics rejection.
The multilayers reflect in the horizontal plane. 
Both substrates have three different multilayer stripes to cover the energy range from 12 to 60\un{keV}, designed and deposited in-house \cite{Morawe2007} (see Table~\ref{tab:dmm-parameters}).

Slope errors are minimized by a ``smart cut'' cooling geometry \cite{Brumund2021a} with white beam illumination over the full length of the multilayer.
This is achieved by setting the horizontal opening of the primary slits to the projected mirror size, $300\un{mm}\cdot \sin(\theta_{\mathrm{DMM}})$, i.e.~between 1\un{mm} (at 3.5\un{mrad}) and 2.1\un{mm} (at 7.4\un{mrad}). 
The resulting power density still allows the use of water-cooling. 

\begin{table}
    \centering
    \begin{tabular}{cccc}
    \toprule
        Stripe Number& \#1& \#2& \#3\\
    \midrule
        Multilayer System & \chem{[Pd(3.50)/C(5.00)]_{20}} & \chem{[W(2.00)/B_{4}C(2.80)]_{30}} & \chem{[W(1.45)/B_{4}C(1.60)]_{70}}\\
        Energy range [keV] & 12 -- 24 & 20 -- 40 & 30 -- 60\\
        Total transmission & 82\% -- 93\% & 66\% -- 87\% & 72\% -- 88\%\\
        d$E/E$ FWHM & 12\% & 6.0\% -- 6.9\% & 2.5\% -- 2.6\%\\
        Incident angles & $\theta$=0.40\textdegree -- 0.20\textdegree & $\theta$=0.40\textdegree -- 0.20\textdegree & $\theta$=0.40\textdegree -- 0.20\textdegree\\
        Stress [GPa] & $-0.78$ & $-1.41$ & $-1.20$\\
        Line force [GPa nm] & 133 & 203 & 256\\
    \bottomrule
    \end{tabular}
    \caption{Design parameters of the multilayer stripes of the DMM.}
    \label{tab:dmm-parameters}
\end{table}


The mechanics of the DMM are derived from an ESRF standard design \cite{Baker2013}.
Both multilayers are mounted on a main plate, which is decoupled from the vacuum vessel by bellows. 
The horizontal position and Bragg angle ($\theta_{\mathrm{DMM}}$) can be set by parallel and antiparallel linear $y$ movements at the upstream and downstream sides of the plate.
Multilayer stripes can be selected by moving the main plate up and down.
The perpendicular distance between the mirror faces is fixed to 3\un{mm}, giving a horizontal offset between the incident white and the doubly-reflected pink beam of $\Delta y = 6\un{mm}\cdot \cos(\theta_{\mathrm{DMM}}) = 6\un{mm}-\epsilon_{\mathrm{DMM}}$, where $\epsilon_{\mathrm{DMM}} \approx 160 \ldots 37\un{nm}$ is negligible compared to the beam size.
The downstream multilayer tracks the diffracted beam on a translation along $x$ on the main plate.
The pitch of the upstream substrate can be fine-tuned with a piezo actuator.
Finally, the roll of the second substrate can be adjusted.


\subsubsection{Block 3: Diagnostics elements} --- Optics block 3 consists of the following elements:

\noindent \paragraph{White beam emergency stop, collimator and second Bremsstrahlung shield:}The white beam stop has a dual function: emergency action to stop the white beam in case of user mistake (when the DMM is out of the beam), and to act as a Bremsstrahlung collimator.

\paragraph{Fixed-width beam limiting aperture:}This limits the horizontal size of the pink beam to 1\un{mm}. 
It also ensures that the power in the pink beam does not exceed 120\un{W}.

\paragraph{Secondary slits:}A new, water-cooled ESRF design, 
able to receive up to 120\un{W} beam power.

\paragraph{Beam intensity monitor:} It consists of a water-cooled filter holder and a 4-quadrant photodiode. Diamond filters coated with different metals (see Table~\ref{tab:filters}) can be selected depending on the energy range to provide a back-scattering signal for the photodiode. The absorbed power is less than 30\un{W}. The back-scattered radiation is measured by a 4-quadrant photodiode with a central aperture, i.e. it does not receive any direct heat load from the beam.
 
\paragraph{Beam profile monitor (beam viewer),} identical to the white-beam viewer installed in Optics Block 1.

\begin{table}
    \centering
    \begin{tabular}{r@{}ll}
    \toprule
        \multicolumn{2}{c}{Position} & Filter \\
    \midrule
        0&.0\un{mm} & (none) \\
        13&.5\un{mm} & 200\un{\mu m} diamond \\
        30&.9\un{mm} & 300\un{\mu m} diamond + 20\un{\mu m} Mo \\
        48&.5\un{mm} & 300\un{\mu m} diamond + 20\un{\mu m} Ag \\
        66&.1\un{mm} & 300\un{\mu m} diamond + 10\un{\mu m} Au \\
    \bottomrule
    \end{tabular}
    \caption{Filter inserts of the beam intensity monitors in Optics Blocks 3, 6 and 7.}
    \label{tab:filters}
\end{table}


\subsubsection{Block 4: Channel-cut crystal monochromator (Si 111)}
--- The horizontally-deflecting, liquid-nitrogen-cooled channel-cut crystal monochromator (CCM) is located at 34\un{m} from the photon source.
It is designed to operate between 12 and 60\un{keV}, i.e.~at Bragg angles $\theta_{\mathrm{CCM}}$ from 9.5\textdegree{} to 1.8\textdegree{}.
The mechanics were supplied by Cinel scientific instruments (Vigonza, Italy). 
The crystal is a monoblock of Si with two (111) surfaces separated by a 4\un{mm} gap, cut and polished in-house. 

The use of the CCM is optional --- for pink beam operations, $\theta_{\mathrm{CCM}}$ is moved to zero and the crystal is translated to $y = -2\un{mm}$, such that the beam passes between the optical surfaces without touching them.
For monochromatic beam operations, the upstream optical surface is moved into the pink beam and $\theta_{\mathrm{CCM}}$ is set to the desired Bragg angle.
The monochromatic beam is then offset by $8\un{mm} \cdot \cos(\theta_{\mathrm{CCM}}) = 8\un{mm}-\epsilon_{\mathrm{CCM}}$ with respect to the pink beam ($\epsilon_{\mathrm{CCM}} \approx 110 \ldots 3.9\un{\mu m}$), i.e.~a total of $\Delta y = 14\un{mm}-\epsilon_{\mathrm{DMM}}-\epsilon_{\mathrm{CCM}}$ with respect to the white beam. 
All downstream instrumentation is designed to be able to work with pink beam ($\Delta y=6\un{mm}-\epsilon_{\mathrm{DMM}}$) and monochromatic beam ($\Delta y=14\un{mm} - \epsilon_{\mathrm{DMM}} - \epsilon_{\mathrm{CCM}}$).

The main motions of the CCM are the horizontal \textit{y}-translation (in/out) , the Bragg angle $\theta_{\mathrm{CCM}}$ and the second crystal pitch correction.


\subsubsection{Block 5: Transfocator} --- The Transfocator device is an ESRF design. It is located at 36\un{m} from the photon source and features 9 pneumatic actuators. Each actuator holds a set of 2D focusing diamond compound refractive lenses (CRLs) from Palm Scientific \cite{Celestre2022}. 
This CRL arrangement of the nine actuators with diamond lenses allows configurations between collimation at 12 keV and focusing onto the sample position at 60 keV. 

In ``pink beam'' mode, the Transfocator will receive a maximum beam power of 120\un{W} from the DMM. Therefore the CRLs  are water-cooled. The useful beam size is approximately $0.75 \times 0.8\un{mm^2}$, therefore the physical apertures of the CRLs are at least this large. For these commercially available diamond lenses with 1\un{mm} material thickness mounted in a 2\un{mm} thick frame, it is implied that the apex radius of curvature is at least 200\un{\mu m}. 
Table ~\ref{tab:transfocator} summarizes the diamond CRLs installed in the Transfocator indicating the number of lenses and the corresponding radius. 

\begin{table}
\centering
\begin{tabular}{cccccccccc}
\toprule
Actuator & \#1 & \#2 & \#3 & \#4 & \#5 & \#6 & \#7 & \#8 & \#9
\\
\midrule
Number of Lenses & 16 & 16 & 16 & 8 & 4 & 2 & 1 & 1 & 1
\\
Radius [\un{\mu m}] & 320 & 320 & 320 & 320 & 320 & 320 & 320 & 640 & 1280
\\
\bottomrule
\end{tabular}
\caption{Diamond lenses in the transfocator.}
\label{tab:transfocator}
\end{table}

\subsubsection{Block 6: Diagnostics module and beam shutter} --- Optics block 6 comprises following elements:
\paragraph{Secondary slits, beam intensity monitor, and beam profile monitor (beam viewer)}: They are identical to those installed in Block 3, and are mounted on a common granite support.

\paragraph{Photon absorber and safety shutter:} They are installed on a separate steel chassis at the downstream end of the Optics Hutch, 38.8\un{m} from the source. 


\subsubsection{Block 7: Diagnostics module} --- At the upstream end of the experimental hutch (EH1) there is the incident beam module with a set of secondary slits, intensity monitor, beam viewer and exit window, similar to previous beam diagnostic blocks 1, 3 and 6. It also comprises a pumping cross for the beam transfer pipe from the optics hutch to the experimental hutch.


\begin{figure}
    \centering
    \includegraphics[width=0.8\linewidth]{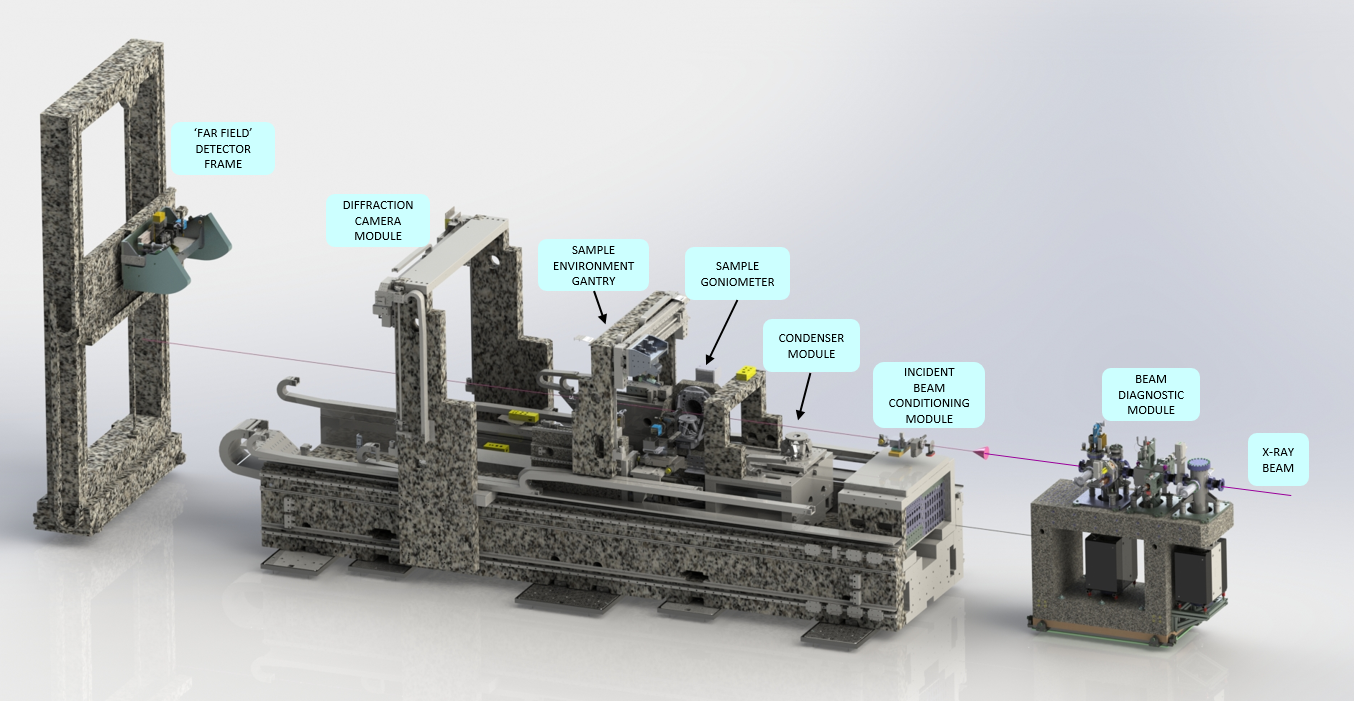}
    \caption{Layout of the end station in EH1.}
    \label{fig:endstation}
\end{figure}

\subsection{End station}

The end station of the beamline is built around the DFXM bench of ID06-HXM \cite{Kutsal2019}. 
This bench was dismantled from the ID06 beamline and was re-installed in the experimental hutch of ID03. 
As shown in Fig.~\ref{fig:endstation}, the setup includes the granite support, detector frame and detectors.
In the support bench there are three main modules: incident beam conditioning module, main block and auxiliary gantry.

\paragraph{Incident beam conditioning module:}

A quadrupole ion chamber is used for beam position and intensity monitoring. It can also be used for feedback to the optics settings in order to stabilize the horizontal beam position. A set of secondary slits, beam viewer and alignment laser is available. The last element of this module is the fast shutter for beam exposure control on the diffraction camera.

\paragraph{Bench main block,} comprising the following modules:

\paragraph{Condenser module:}
This module is the final beam conditioning optics upstream of the sample and remains unchanged from the previous installation. See \cite{Kutsal2019}. The most typical use is to focus the beam to a horizontal line with height (at 17\un{keV}) of $\approx 500\un{nm}$, using 58 1D Be lenses $R=100\un{\mu m}$ (RXOPTICS, Monschau, Germany).

\paragraph{Sample environment gantry:}
A fixed gantry is available there for the installation of sample environments such as a radiation furnace, a liquid nitrogen cryostream, a hot air blower or other devices.

\paragraph{Sample goniometer:}

\begin{figure}[ht]
    \centering
    \vspace{-3mm}
    \includegraphics[width=0.85\linewidth]{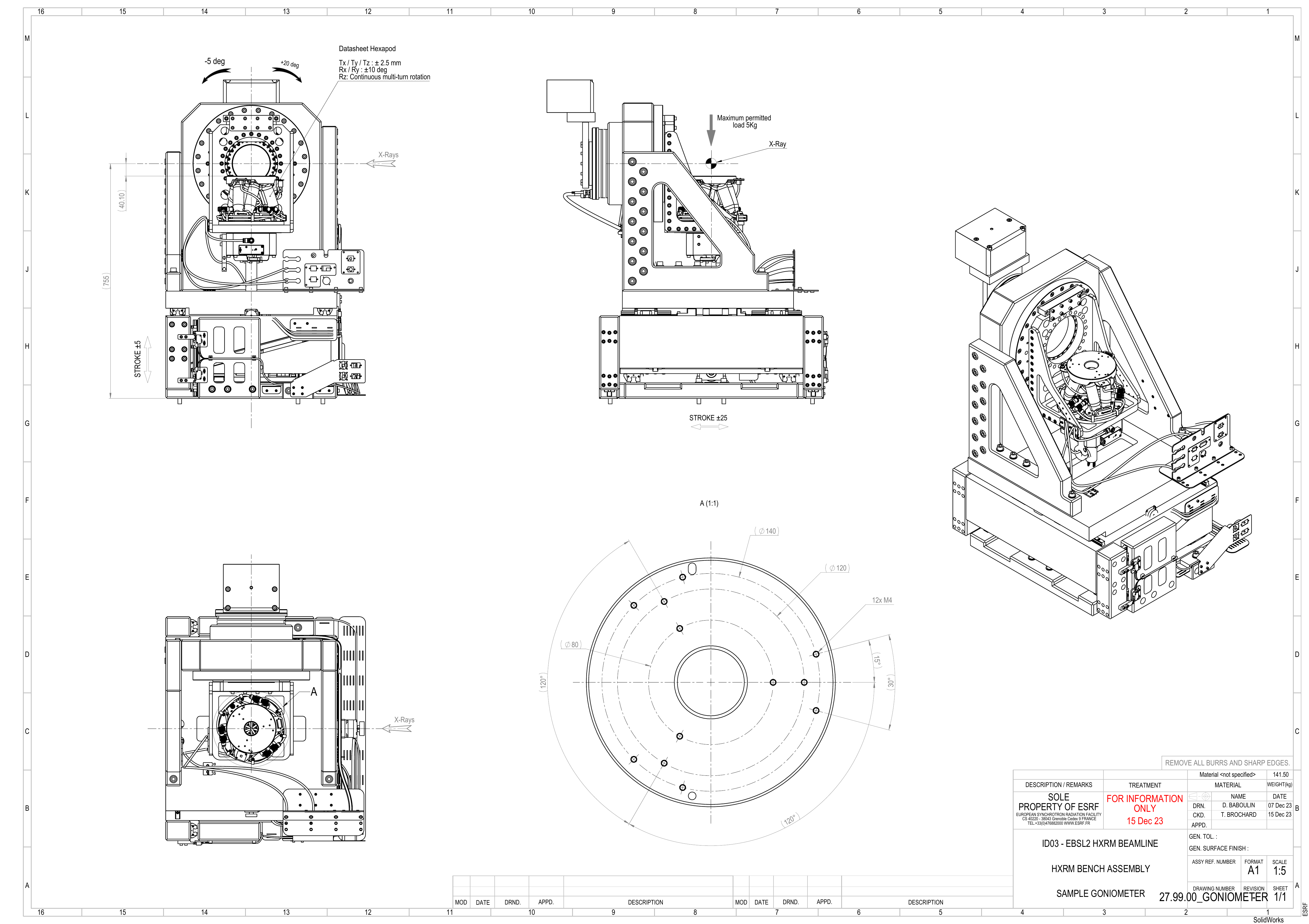}
    \caption{Interface of the goniometer for mounting samples and sample environments.}
    \label{fig:goniometer}
\end{figure}

The goniometer has been upgraded from the previous installation on ID06-HXM \cite{Kutsal2019}.
The new goniometer, constructed by LAB motion systems, is an exact implementation of the geometry described in \citeasnoun{Poulsen2017}. 
It is optimized for topo-tomo type experiments in the vertical diffraction plane \cite{Ludwig2009}.
Briefly, the goniometer offers 4 sample rotations.
From the inside (closest to the sample) out, we have: 

\begin{itemize}

\item Sample translations \texttt{samx}, \texttt{samy} and \texttt{samz}, travel range $\pm 5\un{mm}$.

\item $\phi$: Rotation about $-\hat{y}$ (pitch), travel range $\pm 10$\textdegree.

\item $\chi$: Rotation about $\hat{x}$ (roll), travel range $\pm 10$\textdegree. $\chi$ and $\phi$ are used to align the reciprocal lattice vector under study with the $\omega$ rotation axis.

\texttt{samx}, \texttt{samy}, \texttt{samz}, $\chi$ and $\phi$ are implemented as a hexapod (model Bora by Symetrie, Nimes, France). The hexapod also offers a rotation about $\hat{z}$, which is not used as it is degenerate with the more precise $\omega$ rotation.

\item $\omega$: Air-bearing rotation about $\hat{z}$ (yaw), travel range unlimited (360\textdegree). This rotation is used for topo-tomo studies.

\item $\mu$: Air-bearing rotation about $-\hat{y}$ (pitch), travel range $-5$\textdegree to $20$\textdegree. This rotation is used to set the Bragg angle.

\item Translations to align the goniometer's center of rotation in the beam, \texttt{diffty} (travel range $-25\un{mm}$ to $5\un{mm}$) and \texttt{difftz} (travel range $\pm 3\un{mm}$).

\end{itemize}

The load capacity of the goniometer is 5\un{kg}, for a load with center-of-mass at the center of rotation.
The interface for mounting samples and user sample environments is shown in Fig.~\ref{fig:goniometer}.

\paragraph{Diffraction camera module:}

The diffraction camera module remains unchanged from the previous installation \cite{Kutsal2019}. 
The camera is still a Frelon 2k CCD with 2:1 fiber taper, with effective pixel size of 47.5\un{\mu m}, giving a field of view of about 
$100 \times 100 \un{mm^2}$.
This camera and its beam stop can be moved to with 180\un{mm} of the sample, or to a parking position below the beam path.

\paragraph{Near field camera:}

The near field camera was upgraded to an Optique Peter (Lentilly, France) ``compact white beam'' microscope, with a pco.edge 5.5 sCMOS camera. The scintillator is 18\un{\mu m} of LSO:Tb on a YbSO substrate \cite{Wollesen2022}
In the standard configuration, the objective is a Mitutoyo $10 \times f0.28$, long working distance objective, modified by Optique Peter with lead glass cap to protect against radiation damage. $5 \times f0.21$ and $2 \times f0.55$ objectives are available on demand.
The pixel size of the pco camera is $6.5\un{\mu}$.

\paragraph{Objective module:}

The objective module is used to position the objective lens and an intensity monitor (ionization chamber) in the diffracted beam. 
This module is unchanged from the previous installation \cite{Kutsal2019}. Most experiments use 87 Be lenses (RXOPTICS, Monschau, Germany) with $R=50\un{\mu m}$ for working energies between 17 and 20\un{keV}. Analternative objective lens manufactured from SU-8 polymer (KIT, Karlsruhe, Germany) is available for working energies between 30 and 35\un{keV} \cite{Marschall2014}. Mounted directly downstream of the objective lens is a small ion chamber to monitor the transmitted intensity, which can be used as alignment aid.


\subsubsection{Far field detector frame} --- The far field detector frame mounts two detection systems on a $y$-$z$ alignment stage such that these can be positioned in the Bragg diffracted beam. The reciprocal space camera is composed of a scintillator screen, video objective, and a Basler camera. It is upgraded from the previous configuration \cite{Kutsal2019}, to a Basler acA2440-73gm, with field of view $49 \times 66\un{mm^2}$, and effective pixel size $41\un{\mu m}$. The high resolution camera is based on an Optique Peter ``Twinmic'' dual objective microscope, with a pco.edge 4.2 BI (back-illuminated) sCMOS camera.
In the standard configuration it is equipped with with two Mitutoyo long working distance objectives ($2\times  f0.055$ and $10 \times f0.28$, modified by Optique Peter). 
Each objective has its own scintillator (50\un{\mu m} Europium-doped Gadolinium Gallium Garnet (GGG:Eu) and 25\un{\mu m} GGG:Eu, respectively) and focusing mechanics. 
Users can switch between the objectives remotely by motorized translation.


\section{Data analysis and software}

An online tool for experiment planning is available \cite{Raeder2023}. 
Beamline operation is automated using BLISS as instrument control software \cite{Guijarro2023}.
For data analysis, we have developed the \textit{darfix} package \cite{Garriga2023}, which is based on the Ewoks workflow framework \cite{EWOKS}.
\textit{darfix} provides a set of accessible tools, available as a GUI as well as in the form of Jupyter notebooks, for fast data preprocessing and data analysis, as well as advanced tools that are continuously being developed to fit the needs of the DFXM user community.


%


\section{Outlook}
The new DFXM beamline at ID03 provides significant advancements, offering a $\approx$ 20-fold increase in photon flux at the sample compared to ID06-HXM, along with enhanced stability and a more versatile goniometer. Despite these advances, the current resolution and contrast of the microscopy images are still limited by aberrations in the Be CRLs, which affect the ability to reach the diffraction limit.  Long-wavelength figure errors degrade resolution \cite{Berujon2020}, which we aim to minimize using correction plates \cite{Seiboth2017}. Short-wavelength errors from grain boundaries and roughness create speckle that reduces contrast, prompting exploration of alternative lens materials, such as single-crystal diamond for 30 keV. Alternatives to CRLs, including multilayer Laue lenses \cite{Murray2019} and zone plates, are also being evaluated for improved imaging quality. Looking forward, continued optimization of the optical setup, including advanced lens technologies, pink beam operation, and the extended energy range will enable higher-precision in-situ and operando measurements, broadening the beamline's application to a wider range of materials.


\section*{Acknowledgements}

We thank the numerous members of the ESRF support groups for their contributions. 
Without their work, this project would have been impossible. We acknowledge G.~Winther, H.~Simons, H.~F.~Poulsen and coworkers for their support and numerous contributions to the project over the last 10 years. C.Y. acknowledges the financial support by the ERC Starting Grant "D-REX" (no 101116911).


\bibliographystyle{iopart-num}
\bibliography{main}

\providecommand{\newblock}{}
\begin{thebibliography}{10}
\expandafter\ifx\csname url\endcsname\relax
  \def\url#1{{\tt #1}}\fi
\expandafter\ifx\csname urlprefix\endcsname\relax\def\urlprefix{URL }\fi
\providecommand{\eprint}[2][]{\url{#2}}

\bibitem{Callister2000}
Callister W~D and Rethwisch D~G 2000 {\em Fundamentals of materials science and engineering\/} vol 471660817 (Wiley London)

\bibitem{Roth2015}
Roth T, Detlefs C, Snigireva I and Snigirev A 2015 {\em Opt. Commun.\/} {\bf 340} 33--38

\bibitem{Simons2015}
Simons H, King A, Ludwig W, Detlefs C, Pantleon W, Schmidt S, St{\"o}hr F, Snigireva I, Snigirev A and Poulsen H~F 2015 {\em Nat. Commun.\/} {\bf 6} 6098

\bibitem{Simons2016}
Simons H, Jakobsen A~C, Ahl S~R, Detlefs C and Poulsen H~F 2016 {\em MRS Bulletin\/} {\bf 41} 454

\bibitem{Kutsal2019}
Kutsal M {\em et~al.\/} 2019 {\em IOP Conf. Series: Mat. Sci. Eng.\/} {\bf 580} 012007

\bibitem{yildirim2020}
Yildirim C, Cook P, Detlefs C, Simons H and Poulsen H~F 2020 {\em MRS Bulletin\/} {\bf 45} 277--282

\bibitem{Poulsen2017}
Poulsen H~F, Jakobsen A~C, Simons H, Ahl S~R, Cook P~K and Detlefs C 2017 {\em J. Appl. Cryst.\/} {\bf 50} 1441

\bibitem{Poulsen2018}
Poulsen H~F, Cook P~K, Leemreize H, Pedersen A~F, Yildirim C, Kutsal M, Jakobsen A~C, Trujillo J~X, Ormstrup J and Detlefs C 2018 {\em J. Appl. Cryst.\/} {\bf 51} 1428--1436

\bibitem{Jakobsen2019}
Jakobsen A~C, Simons H, Ludwig W, Yildirim C, Leemreize H, Porz L, Detlefs C and Poulsen H~F 2019 {\em J. Appl. Cryst.\/} {\bf 52} 122

\bibitem{yildirim2023extensive}
Yildirim C, Poulsen H~F, Winther G, Detlefs C, Huang P~H and Dresselhaus-Marais L~E 2023 {\em Scientific Reports\/} {\bf 13} 3834

\bibitem{Snigirev1996}
Snigirev A, Kohn V~G, Snigireva I~I and Lengeler B 1996 {\em Nature\/} {\bf 384} 49

\bibitem{Stoehr2015}
St{\"o}hr F, Wright J, Simons H, Michael-Lindhard J, H{\"u}bner J, Jensen F, Hansen O and Poulsen H~F 2015 {\em J. Micromech. Microeng.\/} {\bf 25} 125013

\bibitem{Tran2021}
Tran~Caliste T~N, Drouin A, Caliste D, Detlefs C and Baruchel J 2021 {\em Applied Sciences\/} {\bf 11} 9054

\bibitem{Garriga2023}
Garriga~Ferrer J, Rodr{\'\i}guez-Lamas R, Payno H, De~Nolf W, Cook P, Sol{\'e}~Jover V~A, Yildirim C and Detlefs C 2023 {\em J. Synchrotron Rad.\/} {\bf 30} 527

\bibitem{Ludwig2001}
Ludwig W, Cloetens P, H{\"a}rtwig J, Baruchel J, Hamelin B and Bastie P 2001 {\em J. Appl. Cryst.\/} {\bf 34} 602--607

\bibitem{Lengeler1999}
Lengeler B, Schroer C, T{\"{u}}mmler J, Benner B, Richwin M, Snigirev A, Snigireva I and Drakopoulos M 1999 {\em J. Synchrotron Rad.\/} {\bf 6} 1153--1167

\bibitem{Falch2016b}
Falch K~V, Casari D, Di~Michiel M, Detlefs C, Snigirev A, Snigireva I, Honkim{\"a}ki V and Mathiesen R~H 2016 {\em J. Mater. Sci.\/} {\bf 52} 3497--3507

\bibitem{Poulsen2001}
Poulsen H~F, Nielsen S~F, Lauridsen E~M, Schmidt S, Suter R~M, Lienert U, Margulies L, Lorentzen T and Juul~Jensen D 2001 {\em J. Appl. Cryst.\/} {\bf 34} 751

\bibitem{Schmidt2004}
Schmidt S, Nielsen S~F, Gundlach C, Margulies L, Huang X and Juul~Jensen D 2004 {\em Science\/} {\bf 305} 229--232

\bibitem{Jakobsen2006}
Jakobsen B 2006 {\em Science\/} {\bf 312} 889--892

\bibitem{Hefferan2012}
Hefferan C~M, Lind J, Li S~F, Lienert U, Rollett A~D and Suter R~M 2012 {\em Acta Mater.\/} {\bf 60} 4311--4318

\bibitem{osti_1476076}
Pokharel R 2018 {\em Overview of High-Energy X-Ray Diffraction Microscopy (HEDM) for Mesoscale Material Characterization in Three-Dimensions\/} (OSTI, 2018)

\bibitem{King2008}
King A, Johnson G, Engelberg D, Ludwig W and Marrow J 2008 {\em Science\/} {\bf 321} 382--385

\bibitem{Ludwig2009}
Ludwig W, Reischig P, King A, Herbig M, Lauridsen E~M, Johnson G, Marrow T~J and Buffiere J~Y 2009 {\em Rev. Sci. Instrum.\/} {\bf 80} 033905

\bibitem{Falch2016a}
Falch K~V, Detlefs C, Di~Michiel M, Snigireva I, Snigirev A and Mathiesen R~H 2016 {\em Appl. Phys. Lett.\/} {\bf 109} 054103

\bibitem{Falch2018a}
Falch K~V, Detlefs C, Snigirev A and Mathiesen R~H 2018 {\em Ultramicroscopy\/} {\bf 184} 1--7

\bibitem{Grunewald2016}
Gr{\"{u}}newald T~A, Rennhofer H, Tack P, Garrevoet J, Wermeille D, Thompson P, Bras W, Vincze L and Lichtenegger H~C 2016 {\em Angew. Chem. Int. Ed.\/} {\bf 55}

\bibitem{Frewein2024}
Frewein M~P~K, Mason J~K, Maier B, C{\"o}lfen H, Burghammer M, Medjahed A~A, Allain M and Gr{\"u}newald T~A 2024 Texture tomography, a versatile framework to study crystalline texture in {3D} arxiv/2404.11195

\bibitem{Carlsen2024}
Carlsen M, Appel C, Hearn W, Olsson M, Menzel A and Liebi M 2024 {\em J. Appl. Cryst.\/} {\bf 57} 986

\bibitem{Morawe2007}
Morawe C, Borel C and Peffen J~C 2007 {\em Advances in X-Ray/EUV Optics and Components II\/} vol 6705 ed Khounsary A~M, Morawe C and Goto S International Society for Optics and Photonics (SPIE) p 670504

\bibitem{Brumund2021a}
Brumund P, Reyes-Herrera J, Detlefs C, Morawe C, del R{\'\i}o M and Chumakov A~I 2021 {\em J. Synchrotron Rad.\/} {\bf 28} 91--103

\bibitem{Baker2013}
Baker R, Barrett R, Clavel C, Dabin Y, Eybert-Berard L, Mairs T, Marion P, Mattenet M, Zhang L, Baboulin D and Guillemin J 2013 {\em J. Phys. Conf. Series\/} {\bf 425} 052015

\bibitem{Celestre2022}
Celestre R, Antipov S, Gomez E, Zinn T, Barrett R and Roth T 2022 {\em J. Synchrotron Rad.\/} {\bf 29} 629--643

\bibitem{Wollesen2022}
Wollesen L, Riva F, Douissard P~A, Pauwels K, Martin T and Dujardin C 2022 {\em J. Mat. Chem. C\/} {\bf 10} 9257--9265

\bibitem{Marschall2014}
Marschall F, Last A, Simon M, Kluge M, Nazmov V, Vogt H, Ogurreck M, Greving I and Mohr J 2014 {\em J. Phys. Conf. Series\/} {\bf 499} 012007

\bibitem{Raeder2023}
R{\ae}der T~M 2023 {\em Journal of Open Source Software\/} {\bf 8} 5177

\bibitem{Guijarro2023}
Guijarro M, Felix L, {De Nolf,} W, Meyer J and G{\"{o}}tz A 2023 {\em Synchrotron Rad. News\/} {\bf 36} 12--19

\bibitem{EWOKS}
De~Nolf W, Payno H, Svensson O and Koumoutsos G 2024 The {ESRF} workflow systen \urlprefix\url{https://doi.org/10.5281/zenodo.6075054}

\bibitem{Berujon2020}
Berujon S, Cojocaru R, Piault P, Celestre R, Roth T, Barrett R and Ziegler E 2020 {\em J. Synchrotron Rad.\/} {\bf 27} 284--292

\bibitem{Seiboth2017}
Seiboth F {\em et~al.\/} 2017 {\em Nat. Commun.\/} {\bf 8}

\bibitem{Murray2019}
Murray K~T, Pedersen A~F, Mohacsi I, Detlefs C, Morgan A~J, Prasciolu M, Yildirim C, Simons H, Jakobsen A~C, Chapman H~N, Poulsen H~F and Bajt S 2019 {\em Opt. Express\/} {\bf 27} 7120

\end{thebibliography}

\end{document}